# ESTIMATION RADIUS REGRESSION IN PLIC-VOF METHOD FOR DROPLET EVAPORATION

Jean NAHED[1*], Joseph DGHEIM[2*]

[*]*Laboratory of Applied Physics (LPA), Group of Mechanical, Thermal & Renewable Energies (GMTER), Lebanese University, Faculty of Sciences, Lebanon*

## ABSTRACT

Navier-Stokes and energy equations of hydrocarbon liquid droplet are solved numerically using a semi-implicit method based on the finite volume scheme (*VOF*). A radius regression estimation technique based on the *PLIC-VOF* method is developed. The determination of the normal vector and the curvature are based on the **J. Nahed et al.** technique [1]. Numerical calculation is performed for stagnant hydrocarbon droplets, in evaporation in natural convection. Good agreement between the numerical results and the experimental ones is observed. The results are presented in the form of radius, area, mass regressions, volumetric mass flow, liquid temperatures, surface temperatures, and vapor temperatures by taking into consideration the effect of the ambient temperature, the radius, and the component variations. Using the time evolution of the color function, our radius regression estimation technique is based on instant detection of liquid volume fraction in each cell of the mesh and it is not affected by the changing of the droplet's shape nor the temperature collecting and synchronization with the method for the evaporation analysis.

**Nomenclature**

| | |
|---|---|
| $a$ | integer number |
| $b$ | real number |
| $C$ | circle |
| $Cp$ | specific heat, J.kg$^{-1}$.K$^{-1}$ |
| $e$ | exponential term |
| $F$ | color function |
| $g$ | terrestrial acceleration, m.s$^{-2}$ |
| $k$ | curvature, m$^{-1}$ |
| $L$ | latent heat, J.kg$^{-1}$ |
| $Max$ | maximum |
| $\dot{m}$ | volumetric mass flow, kg.m$^{-3}$.s$^{-1}$ |
| $N$ | total number of cells |
| $n$ | normal |
| $\vec{n}$ | normal vector |
| $\|\vec{n}\|$ | norm of the normal vector |
| $P$ | pressure, atm |
| $r$ | sphere radius, m |
| $R$ | curvature radius, m |
| $S$ | surface dimension, m$^2$ |
| $t$ | time, s |
| $T$ | temperature, K |
| $TL$ | liquid temperature, K |
| $TS$ | surface temperature, K |
| $TV$ | vapor temperature, K |
| $u$ | velocity component in ordinate direction, m.s$^{-1}$ |
| $v$ | velocity component in coordinate direction, m.s$^{-1}$ |
| $w$ | real number |
| $x$ | x axis, m |
| $y$ | y axis, m |

*Greek letters*

| | |
|---|---|
| $\alpha$ | thermal diffusivity, m$^2$.s$^{-1}$ |
| $\beta$ | expansion number |
| $\partial$ | differential |
| $\Delta$ | difference |
| $\eta$ | dimensionless radius |
| $\lambda$ | thermal conductivity, W.m$^{-1}$.K$^{-1}$ |
| $\mu$ | dynamic viscosity, Kg.m$^{-1}$.s$^{-1}$ |
| $\vartheta$ | kinematic viscosity, m$^2$.s$^{-1}$ |
| $\rho$ | density, kg.m$^{-3}$ |
| $\Sigma$ | summation |
| $\sigma$ | surface tension, N.m$^{-1}$ |
| $\theta$ | polar angle, ° |

*Subscripts*

| | |
|---|---|
| *boil* | boiling |
| *g* | gas |
| *i* | incremental step in ordinate direction |
| *j* | incremental step in coordinate direction |
| *k* | phases index |
| *l* | liquid |
| *s* | surface |
| $\eta$ | radius axis |
| $\theta$ | polar axis |
| *0* | initial |
| $\infty$ | ambient medium |

*Superscripts*

| | |
|---|---|
| *c* | real number |
| *n* | iteration number |
| *0* | initial |

## I. Introduction

Studies to improve the internal combustion in engines, turbo machines, and chemical products are based on research in liquid droplets for years. Recently, the progress of numerical studies in liquid droplet evaporation has influenced the development of new numerical techniques. The two-phase flow phenomenon, existing in the evaporation of the hydrocarbon liquid droplet is remaining an important field for the research committee. Despite the development of new numerical techniques, the two-phase flow phenomenon is still a difficult field in researches [2-3]. Many authors [4-5] have solved this problem and developed new methods. The main procedure of these methods is to follow the time evolution of the radius regression and the transportation of the interface. Most studies focused on the evaporation of isolated droplets are considered by many authors such as **Vysokomornaya et al.** [6]. The essential phenomenon involved in actual fuel injection and atomization processes is the evaporation of single isolated fuel droplets. The evaporation phenomenon of isolated fuel droplets is based on the fundamental processes of heat and mass transfer, and the information derived can be applied to various fuel spray patterns and combustor geometries. Although great improvement has been made in developing new powerful numerical methods, information derived from isolated fuel droplet researches still has a key role in the success of numerical models for spray evaporation [7]. The evaporation of fuel droplets in internal combustion engines has a powerful influence on emissions of the pollutant particles, ignition delays, and overall combustion kind. When fuels droplets are injected in the combustion chamber into a high-temperature environment, the droplets are heated, evaporate and the fuel vapor burns to deliver energy for propulsion [8]. Spray evaporation has been thoroughly studied in the last decades, theoretically, experimentally, and numerically [9-10]. But, the complexity of such phenomenon, a single droplet evaporating in a high-temperature ambient medium, which represents an ideal image of the physical phenomenon involved in the diluted regions of a spray, is a perfect model to better understanding the evaporation dynamics. Heating and evaporation types of single droplets are explained in detail in **Sirignano** [11] and **Sazhin** [12]. Many experimental investigations have been applied to study the evaporation of isolated droplets for different fuels and under several ambient medium conditions [13-14]. In the meantime, many numerical investigations have also been applied and many theoretical estimations have been verified in comparison with experimental results since spray simulation results are strongly influenced by the droplet evaporation models adopted. Despite the possessions and basis of information on isolated droplets' evaporation achieved in the last decades, research on this theme is still ongoing due to the massive influence of gas and diesel engines and their performance on the emissions' characteristics.

---
[1] Author, Email address: naejdehan@hotmail.com
[2] Corresponding author, Email address: jdgheim@ul.edu.lb,
Lebanese university, Fanar, Jdeidet El-Metn, Liban
P.B.: 90656
tel./fax: (961) 1 68 15 53

Natural convection has a significant influence on the modification of heat and mass transfer rates between the evaporating liquid phase and the surrounding gaseous phase. The circumstances found in a combustor chamber both increase and decrease in importance the natural convection due to the high-pressure and high-temperature environment and small droplet sizes, respectively. Though, there are a few quantitative experimental facts to confirm that small fuel droplets, evaporating at elevated pressure and temperature are unaffected by natural convection. In phenomena where natural convection is influential, droplet size becomes a significant factor in the application of large sphere information to the practical spray evaporation phenomenon, as the mean droplet size of most practical systems is less than 100 μm. Many studies continue to examine droplets with initial diameters on the order of 1000 μm [15]. On the other hand, droplet evaporation under normal gravity conditions is affected by natural convection. This scenario enhances energy and mass transfers between the evaporating liquid phase and the surrounding gaseous phase. Hence, depending on the ambient medium conditions, incorporating these effects might become essential to correctly predict droplet evaporation rate [16]. Therefore, in this paper, the phenomena of natural convection and normal gravity are performed.

Energy deficiency and environmental protection are scientific, research, and industrial challenges for the future development of the internal combustion engine. Thus, researchers find petroleum alternative fuels get much attention. Methanol and Ethanol as oxygenated renewable biomass fuels are accepted clean alternative fuels [17-18]. Yet, when biomass fuels are mixed with other fuel kinds, due to the big differences of boiling points, density, latent heat of vaporization, surface tension, and viscosity, the physical properties of the blended fuel will change and thus affect the evaporation and the combustion inside the internal combustor chamber and the performance and emissions of the engine [19]. Hence, in this study, the oxygenated biomass fuels are taken into consideration.

Several researchers like **Bhattacharya et al.** [20] and **Dgheim et al.** [21, 22] performed many studies on regression of droplet radius, evolution of droplet surface, liquid and vapor phase temperatures. These authors studied numerically and experimentally the evaporation and combustion of the fuel droplets in natural and forced convection, in normal gravity. In their investigations, the radius regression is verified using the $d^2$ law. Early numerical and experimental studies concerning the evaporation of isolated droplets drive to the fundamental $d^2$ law. This last shows, most notably, that the regression in droplet area varies linearly with time. To analyze the evaporation process with $d^2$, it is necessary to ensure that the temperature collecting must react fast and accurately and can synchronize with $d^2$. The shape of the droplet keeps changing during the evaporation process, which brings many difficulties to obtain the $d^2$ of the droplet. Various methods are applied by different researchers. **Gan and Qiao** [23] counted the number of pixels occupied in the droplet by setting a threshold, solved the droplet diameter with Matlab. **Morin et al.** [24] used the binary method to extract the profile of the droplet and used the integral method to obtain the surface area of the droplet and then the droplet diameter. **Yang and Wong** [25] obtained the droplet boundary with image processing software and then calculated the diameter by assuming that the droplet was ellipsoid.

The primary purpose of this paper is a new radius regression technique based on instant detection of liquid volume fraction in each cell of the mesh using the evaporation characteristics of liquid droplets to prove the accuracy of our model. This estimation method is a new approach to the radius regression and a progression of the last developed one [1] and it is not affected by the changing of the droplet's shape nor the temperature collecting and synchronization with the method for evaporation analysis.

The paper is organized as follows. In section 2, the mathematical models adopted and a description of the mathematical formulations of our technique are presented. Numerical and experimental details are stated and the results are discussed in section 3. Finally, section 4 summarizes the main findings and conclusions.

## II. Mathematical modeling

Heat and fluid flow transfers' equations in liquid droplet evaporation are solved by taking into consideration our radius regression estimation method. The normal vector and the curvature are calculated using **Nahed et al.** [1] technique.

### II.1 Mathematical equations

Our model based on the heat and fluid flow transfers' equations used variable change in order to transform the equations from spherical coordinates to Cartesian coordinates:

$$(t, r) \to \left(t, \eta = \frac{r}{r_s(t)}\right) \quad (1)$$

$$\frac{\partial}{\partial t}\Big|_{t,r} = \frac{\partial}{\partial t} - \frac{\eta}{r_s}\frac{dr_s}{dt}\frac{\partial}{\partial \eta}\Big|_{t,r} \quad (2)$$

$$\frac{\partial}{\partial r}\Big|_{t,r} = \frac{1}{r_s}\frac{\partial}{\partial \eta}\Big|_{t,r} \quad (3)$$

$$\frac{\partial^2}{\partial r^2}\Big|_{t,r} = \frac{1}{r_s^2}\frac{\partial^2}{\partial \eta^2}\Big|_{t,r} \quad (4)$$

Then, our mathematical equations become as the following:
Continuity equation:

$$\frac{1}{r_s}\frac{\partial u_k}{\partial \eta} + \frac{1}{\eta r_s}\frac{\partial v_k}{\partial \theta} + \frac{2u_k}{\eta r_s} + \frac{v_k \cos\theta}{\eta r_s \sin\theta} = 0 \quad k = l, g \quad (5)$$

Momentum equation:

$$\frac{\partial u_k}{\partial t} = +\frac{\vartheta_k}{r_s^2}\left(\frac{2}{\eta} + \frac{\eta r_s}{\vartheta_k}\frac{dr_s}{dt} - \frac{u_k r_s}{\vartheta_k} + \frac{1}{\mu_k}\frac{\partial \mu_k}{\partial \eta}\right)\frac{\partial u_k}{\partial \eta} + \frac{\vartheta_k}{r_s^2}\frac{\partial^2 u_k}{\partial \eta^2} + \frac{\vartheta_k}{r_s^2}\left[\frac{\cos\theta}{\eta^2 \sin\theta} - \frac{v_k r_s}{\eta \vartheta_k}\right]\frac{\partial u_k}{\partial \theta} + \frac{\vartheta_k}{\eta^2 r_s^2}\frac{\partial^2 u_k}{\partial \theta^2} - g\beta_T(T_k - T_\infty)\sin\theta - \frac{\sigma k}{\rho r_s} + \frac{\sigma k n}{\rho r_s} \quad k = l, g \quad (6)$$

Energy equation:

$$\frac{\partial T_k}{\partial t} = +\frac{\alpha_k}{r_s^2}\left(\frac{2}{\eta} + \frac{\eta r_s}{\alpha_k}\frac{dr_s}{dt} - \frac{u_k r_s}{\alpha_k} + \frac{1}{\lambda_k}\frac{\partial \lambda_k}{\partial \eta}\right)\frac{\partial T_k}{\partial \eta} + \frac{\alpha_k}{r_s^2}\frac{\partial^2 T_k}{\partial \eta^2} + \frac{\alpha_k}{r_s^2}\left[\frac{\cos\theta}{\eta^2 \sin\theta} - \frac{v_k r_s}{\eta \alpha_k}\right]\frac{\partial T_k}{\partial \theta} + \frac{\alpha_k}{\eta^2 r_s^2}\frac{\partial^2 T_k}{\partial \theta^2} \quad (7)$$

Where $k=l$ or $g$
Evolution equation:

$$\frac{\partial F}{\partial t} - \frac{\eta}{r_s}\frac{dr_s}{dt}\frac{\partial F}{\partial \eta} + \frac{1}{r_s}\frac{\partial u_k F}{\partial \eta} + \frac{1}{\eta r_s}\frac{\partial v_k F}{\partial \theta} = 0 \quad (8)$$

The thermo-physical properties of the evaporated droplet are given by:

$$\rho = \rho_l F + \rho_g(1 - F) \quad (9)$$
$$\mu = \mu_l F + \mu_g(1 - F) \quad (10)$$
$$\lambda = \lambda_l F + \lambda_g(1 - F) \quad (11)$$
$$C_p = C_{pl} F + C_{pg}(1 - F) \quad (12)$$
$$\alpha = \alpha_l F + \alpha_g(1 - F) \quad (13)$$
$$\vartheta = \vartheta_l F + \vartheta_g(1 - F) \quad (14)$$

The curvature is determined as:

$$k_{i,j} = \frac{1}{\|\vec{n}\|.e^{r_{i,j}}} \quad (15)$$

Where $r_{i,j} = \sqrt{\left(|F_{i+1,j} - F_{i,j}|r_s \sin\theta\right)^2 + \left(|F_{i,j+1} - F_{i,j}|r_s \cos\theta\right)^2}$

The normal vector is the following:

$$\vec{n} = -\frac{rs}{2a\pi}\int_0^{2\pi}\int_0^1 k \, d\eta \, d\theta \quad (16)$$

The volumetric mass flow is determined as the following:

$$\dot{m} = 3\frac{\rho_l}{r_s}\frac{dr_s}{dt} \quad (17)$$

These equations are discretized using the semi-implicit finite volume scheme.
A study of the calculus stability for a given space step $\Delta\eta$ leads to a time step $\Delta t = 0.01$ s. The value of these steps leads to a scheme of 30x30. The system of the algebraic equations is solved using Zaleski Algorithm.

### II.2 Radius regression estimation

In the *PLIC-VOF* method, our radius regression estimation consists of the calculation of the liquid volume fraction, $F_{i,j}$, in each cell of two-dimensional mesh, as presented in (Fig 1).

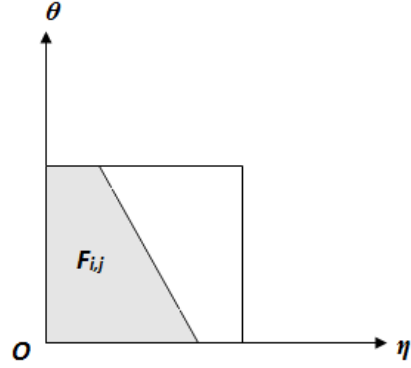

**Fig 1**: Liquid fraction of a local cell for two-dimensional mesh.

Then, the sum of the liquid volume fractions all over the two-dimensional mesh is the total volume of the liquid droplet. Thus, the dimensionless ratio $\frac{\Sigma_{i,j}^N F_{i,j}^{(n)}}{\Sigma_{i,j}^N F_{i,j}^{(0)}}$, where the superscript $n$ is the number of iterations, is the liquid volume fraction evaporated in $n$ iterations, as showed in (Fig 2).

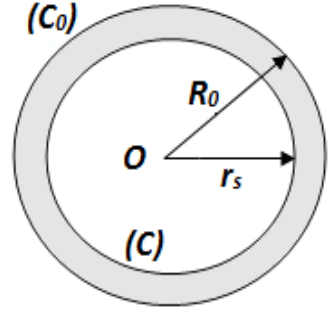

**Fig 2**: Total liquid fraction evaporated in $n$ iterations for two-dimensional mesh.

Our calculation consists of two steps: The first step is the determination of the radius $r_s$ and the second one is the optimization of our radius calculation using the least square method.
Firstly, surfaces $S$ and $S_0$ of the circles $(C)$ and $(C_0)$ are used to calculate the radius $r_s$ of the liquid droplet as the following:

$$S = \pi r_s^2 \quad (18)$$
$$S_0 = \pi R_0^2 \quad (19)$$

Where $R_0$ and $r_s$ are the initial and calculated droplet radii respectively.

The dimensionless ratio $\frac{S}{S_0}$ is the fraction indicated in (Fig 2). Secondly, the radius regression calculation is optimized by using the least square method [26] and our validated experimental results [2] as presented in figure 3. One can find the following relationship-with a standard error of 0.006:

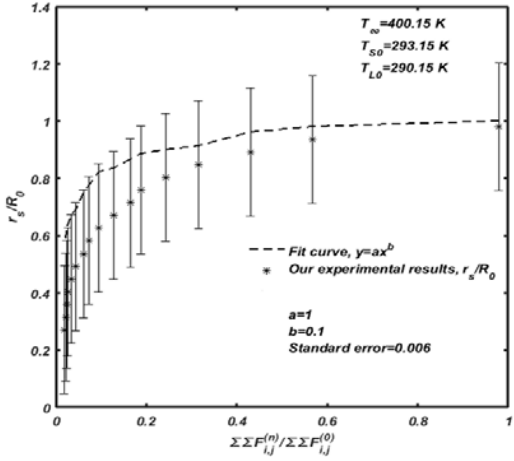

**Fig 3:** Comparison of experimental and numerical results for radius regression ratio.

$$r_s = \left(\frac{\sum_{i,j}^N F_{i,j}^{(n)}}{\sum_{i,j}^N F_{i,j}^{(0)}}\right)^{0.1} R_0 \qquad (20)$$

The radius square is given by:

$$r_s^2 = \left(\frac{\sum_{i,j}^N F_{i,j}^{(n)}}{\sum_{i,j}^N F_{i,j}^{(0)}}\right)^{0.2} R_0^2 \qquad (21)$$

The radius square difference is written as the following:

$$\Delta r_s^2 = r_{s1}^2 - r_{s2}^2 \qquad (22)$$

Where $r_{s1}$ and $r_{s2}$ are the radii at different conditions. We used the mathematical criterion $\Delta r_s^2$ to describe the distribution of the $r_s^2$ variations compared to fixed values of $r_s^2$ at fixed conditions, such as ambient temperature, liquid component and others, and to show the difference level when the conditions change.

According to our model, the calculation of the radius $r_s$ is achieved through the instant detection of the liquid volume fraction in each cell of a two-dimensional mesh.

### III. Results & Discussion

Our numerical calculation is performed for different droplet radii, different ambient temperatures and several types of hydrocarbon liquid droplets. The results of our model are presented under the form of linear and surface profiles. Our experimental set-up is presented in our previous work [2].

### III.1 Model validation and linear results

The accuracy of our numerical method is proved using several linear results. These results are performed using a mesh of 30x30.

In evaporation, in natural convection, the heptanes droplet area regression is presented numerically and experimentally (Fig 4)

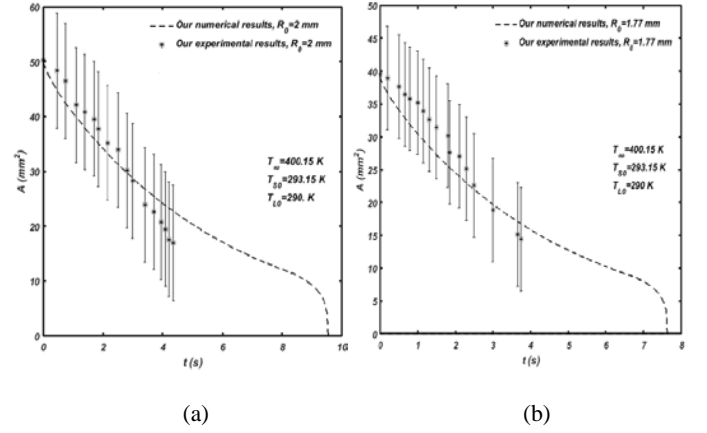

(a)                    (b)

**Fig 4:** Comparison of experimental and numerical results for area regression (a) initial radius $R_0 = 2\ mm$, (b) initial radius $R_0 = 1.77\ mm$.

for two different droplet radii. Good qualitative and quantitative agreements are observed between experimental and numerical results. The relative error is 2.85 % and 1.96 % respectively. Also, the evaporation process accelerates the decreasing of the initial radius and shortens its time length. According to equation (27), the regression of the square radius and the mass of heptane liquid droplets for different initial radii, presented in figures 5 and 6, decrease with time. These results show that the smaller is the droplet; the shorter is the time of evaporation.

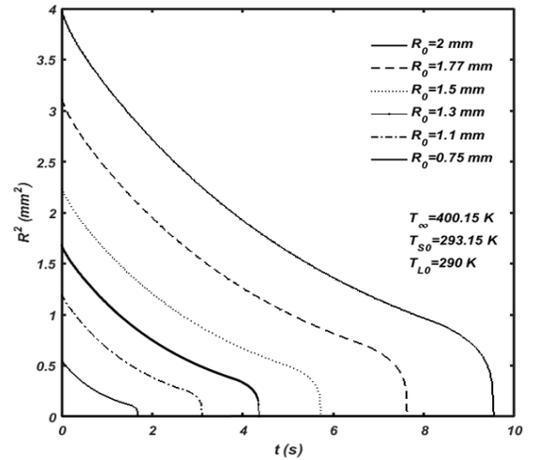

**Fig 5:** Radius square regression versus time for different initial radii.

The inflection and the straight fall of the curve show the end of the evaporation process and calculation. The mass regression of the heptane droplets for different initial radii indicates that the time of mass transfer is shorter when the droplet mass is smaller.

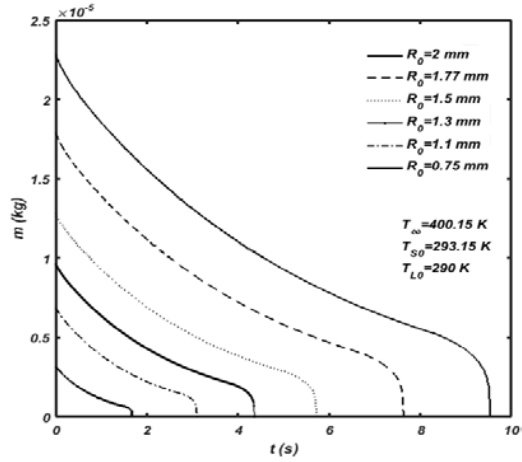

**Fig 6:** Mass regression versus time for different initial radii.

Furthermore, the mass flow which is in direct relation with the radius regression (Eq. 17), presented in (Fig 7) and based on the same mass sizes and conditions of the previous heptane droplets, proves that the mass transfer increases during the first stage of the evaporation, when the surface temperature increases to get its saturation value under the boiling point, and remains constant to the end of the evaporation phenomenon when the mass transfer begins decreasing.

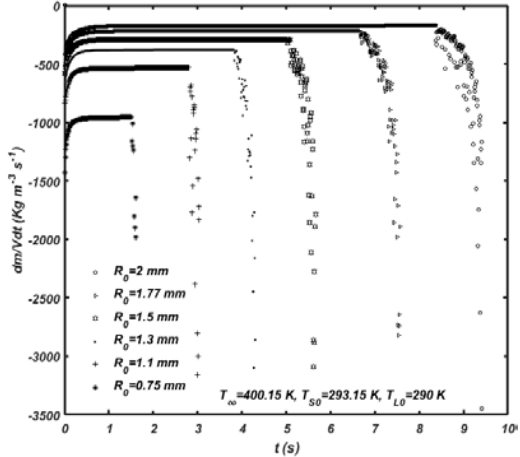

**Fig 7:** Volumetric mass flow evolution versus time for different initial radii.

The radius square difference, $\Delta r_s^2$, shows a clear modification in the evaporation rate when the ambient temperature changes, as presented in figures 8 and 9 respectively.

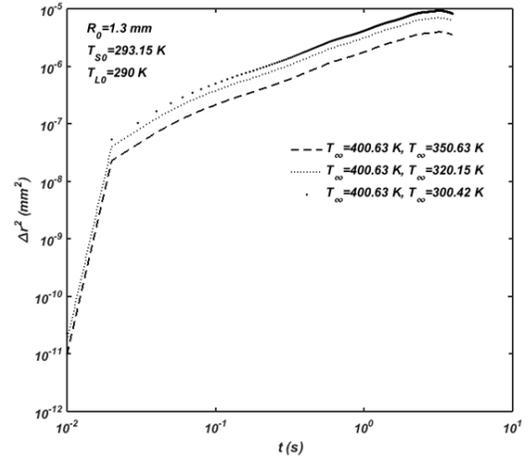

**Fig 8:** Difference of radius square variation versus time for different ambient temperatures.

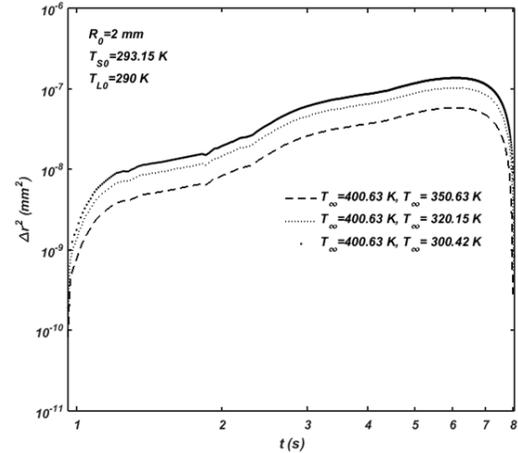

**Fig 9:** Difference of radius square variation versus time for different ambient temperatures

The heptane liquid droplet at $T_\infty = 400.63\ K$ is compared to other heptane liquid droplets at different ambient temperatures. The time of evaporation is shorter as the ambient temperature is greater. Moreover, the evaporation rate is modified with the variation of the droplet's physical properties, as shown in (Fig 10). $\Delta r_s^2$ indicates an obvious difference between the radius square regression of liquid droplets for different physical properties. The hexane liquid droplet is compared to other hydrocarbon liquid droplets under the same conditions. To prove the accuracy of our method, another comparison between the experimental and the numerical results of the evolution of the heptane surface temperature in evaporation in natural convection is realized. Figure 11 indicates that our numerical results, for the surface temperature, show good agreement comparing to those of the experimental ones with a relative error of 0.098%. This last means that our numerical method reflects well the experimental one.

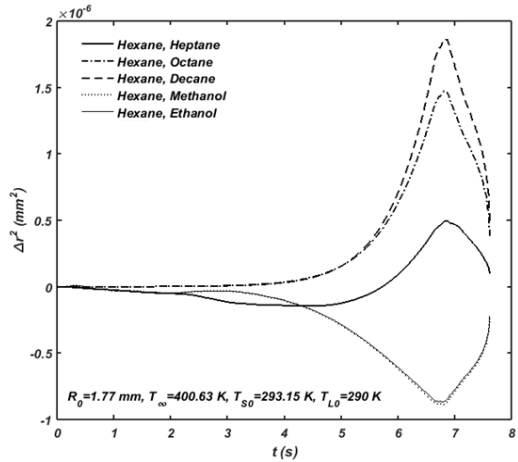

**Fig 10:** Difference of radius square variation versus time for different components.

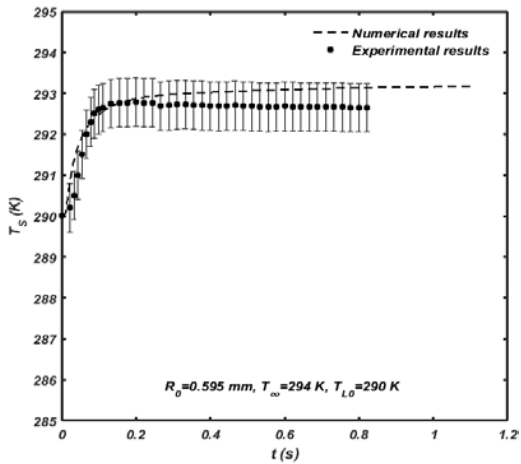

**Fig 11:** Comparison of experimental and numerical results for surface temperature.

The surface temperatures of a decane liquid droplet for different ambient temperatures are computed versus time and radius square (Figures 12.a and 13.a) to more verify the adequacy of our numerical method. Increasing in temperature gradient produces increasing in the surface temperature. As a result, the heating time of the surface is shorter which is visible in both figures. Furthermore, whatever the temperature gradient is high, the surface temperature no longer exceeds the boiling temperature, $T_{boil} = 447.3\ K$, and remains constant till the total evaporation of the droplet which proves the accuracy of our numerical technique, the boundary condition on the surface, and the good preservation of the heat transfer.

Under the same conditions, the temperatures of liquid and vapor phases of the decane liquid droplet for different ambient temperatures are computed versus time and radius square in figures 12.b; 13.b and 12.c; 13.c respectively. As well, increasing the temperature gradient produces increasing in the liquid and vapor temperatures, as is visible in these figures. As a result, the heating time of the liquid phase is shorter. The liquid and vapor temperatures rise to get their limits. The limit values of our temperatures are considered optimal because they reach the value of the surface temperatures for the liquid phase, represented in figure 12.b and figure 13.b, and the ambient temperatures for the vapor phase, represented in figure 12.c and figure 13.c. As well, the recent figures show that the two-phase flows of the heat transfer, versus time and radius square, are well-considered.

To better prove the accuracy of our numerical method, other linear results concerning the temperature evolution are realized. Using different types of hydrocarbon liquid droplets in evaporation in natural convection, and under the same conditions, the surface temperature, and the liquid and vapor temperatures are computed versus time and radius square in figure 14 and 15 respectively. Surface temperatures of each type of hydrocarbons, represented in figures 14.a and 15.a, no longer exceed their related boiling temperature. Besides, they reach their highest values relative to the ambient temperature and they remain constant till the total evaporation of the droplet. Moreover, the evaporation time of the ethanol droplet is shorter than the other hydrocarbon droplets because it is the most volatile. Its surface temperature is higher than the one of the hexane droplet and reaches the boiling temperature, $T_{boil} = 351.5\ K$, quicker than the hexane's surface temperature. The heat transfer of all hydrocarbon liquid droplets is well-considered and compatible with their volatile degree. Figures 14.b and 15.b demonstrate that the liquid temperature of each type of hydrocarbon liquid droplet reaches its limit value, the surface temperature. This limit value is observed depending on the hydrocarbon's physical property. This last affects the temperature gradient and the heat transfer of each droplet. Thus, it reflects the variation of the liquid temperatures. Figures 14.c and 15.c represent the evolution of the vapor temperatures. All hydrocarbon vapor temperatures start from the surface temperature and reach their limit value, the ambient temperature. These figures prove that the heat transfer in liquid-vapor phases, for each liquid droplet, is well considered.

### III.2 Model validation & results of surface profiles

The results of the surface profiles are realized by studying the effect of temperature gradient on the liquid droplet's free surface, the heat and mass transfer, and the evaporation of the hydrocarbon droplets of various initial radii, initial temperatures, and components.

Figure 16 shows the isothermal sequence, calculated by the energy equation, inside the liquid and vapor phases and at the liquid surface for an octane liquid droplet in evaporation in natural convection. The sequence represents the gradual circular spreading of vapor in the surrounding area with time. Therefore, the increase in temperatures of liquid and vapor phases is seen. Thus, the heat transfer is observed and well-considered. At the initial time, t=0s, figure 16.A represents the initial temperatures of the liquid droplet and its ambient medium. At t=0.1s of figure 16.B, the temperature gradient activates the heat transfer, and thus the temperature of the liquid droplet increases. Then, the evaporation phenomenon of the liquid phase appears and a layer of vapor is observed around the liquid droplet. After 0.5s, the heat transfers to the two phases, the liquid and the vapor phases (figure 16.C) increase their temperatures to get their limit values. Also, the evaporation phenomenon is observed and the vapor layer is thicker.

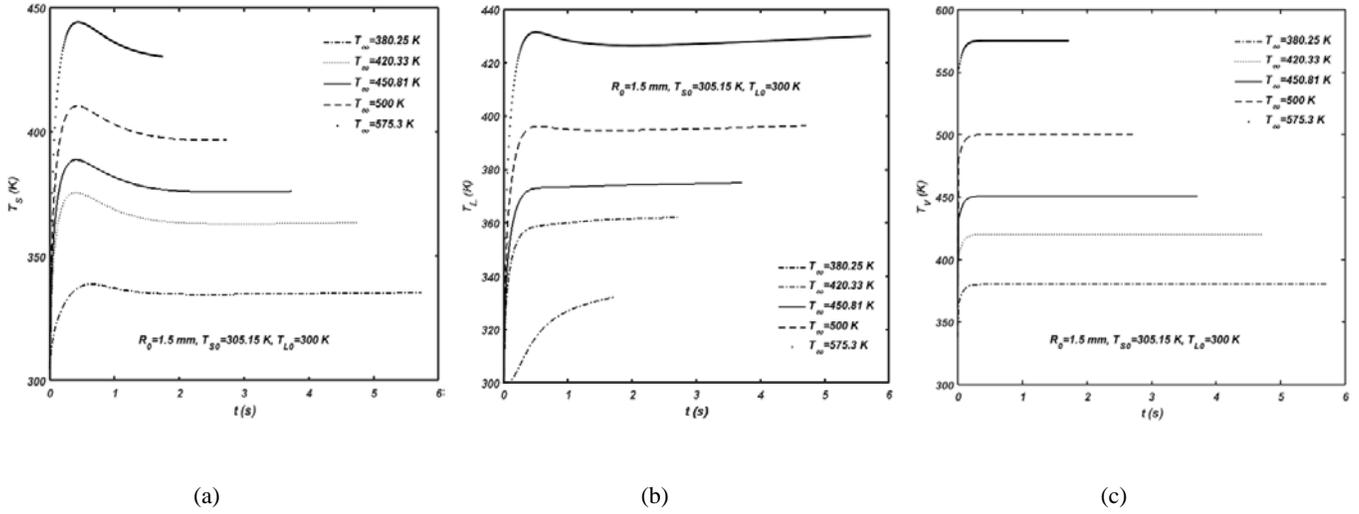

(a)          (b)          (c)

**Fig 12:** (a) Surface temperature, (b) Liquid temperature and (c) Vapor temperature versus time for different ambient temperatures

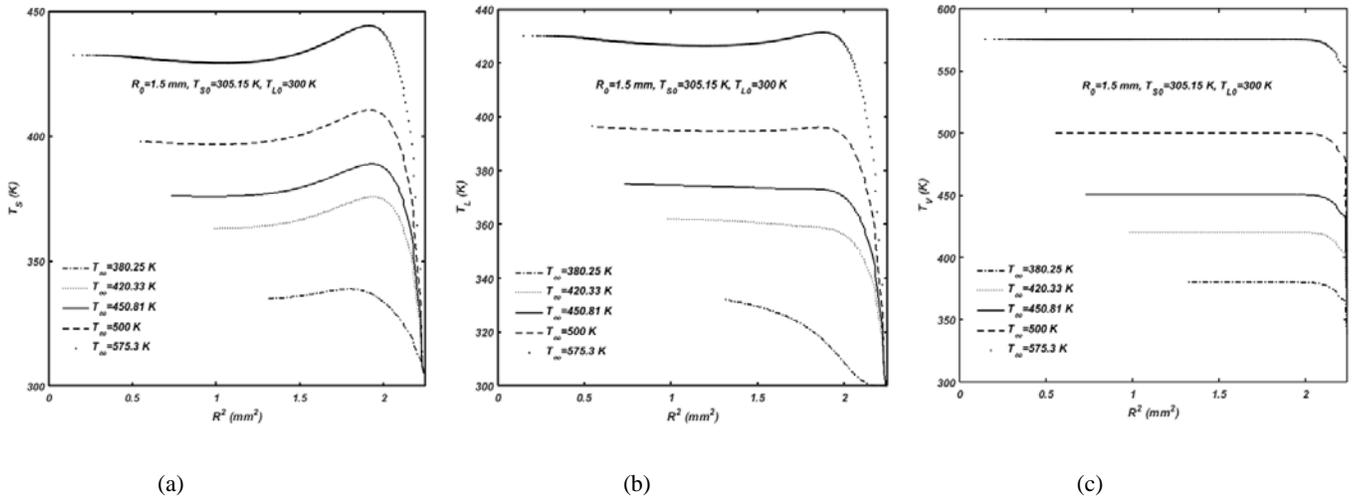

(a)          (b)          (c)

**Fig 13:** (a) Surface temperature, (b) Liquid temperature and (c) Vapor temperature versus radius square for different ambient temperatures

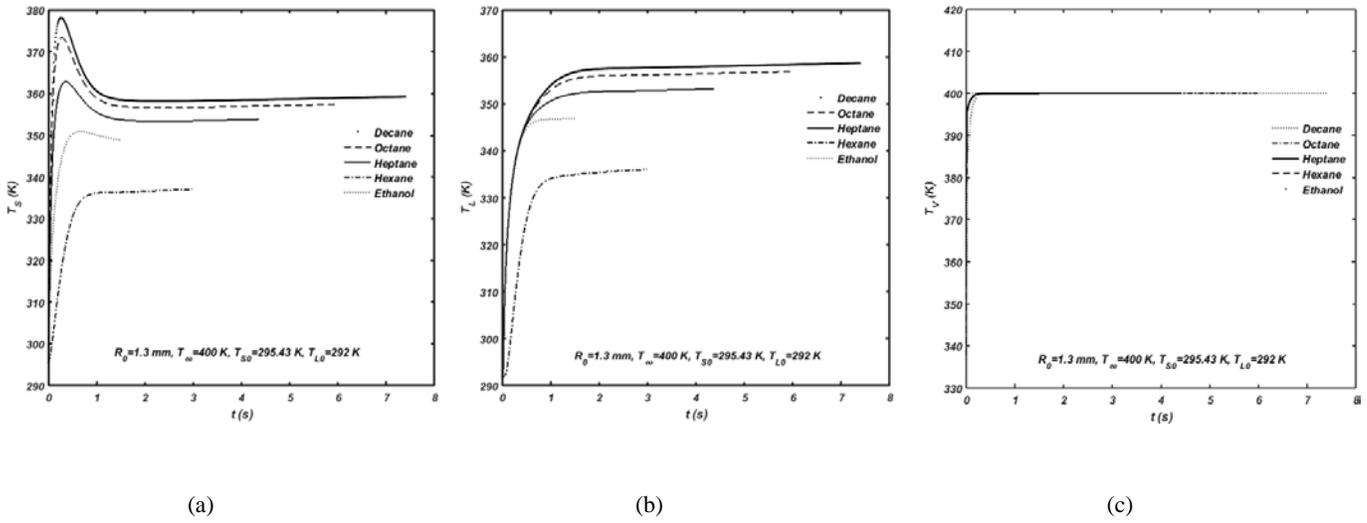

(a) (b) (c)

**Fig 14:** (a) Surface temperature, (b) Liquid temperature and (c) Vapor temperature versus time for different components.

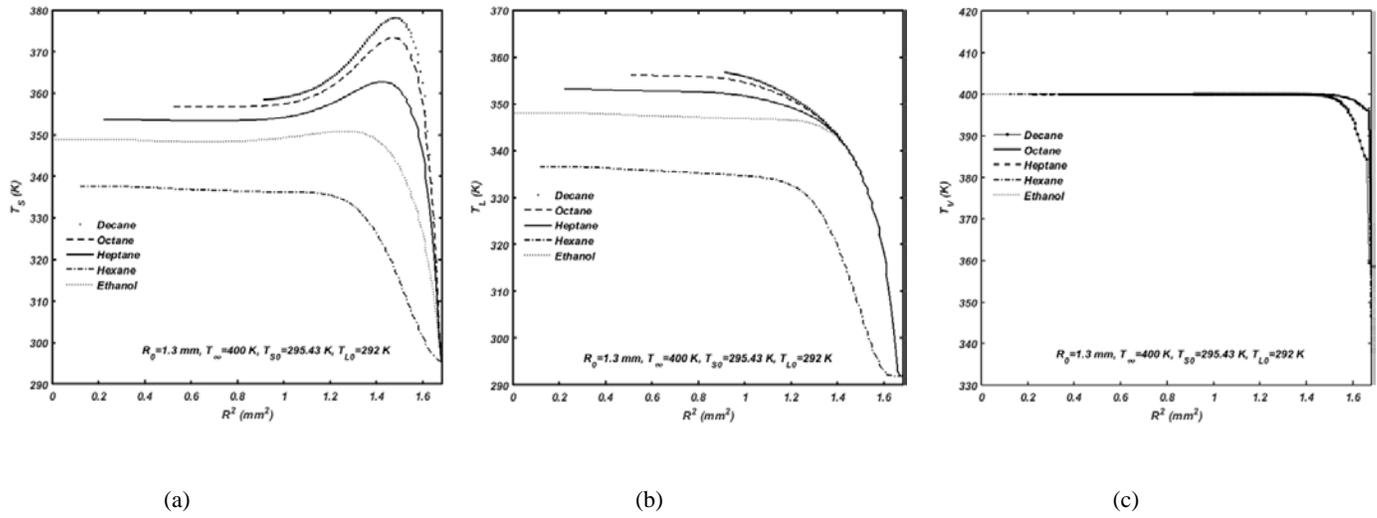

(a) (b) (c)

**Fig 15:** (a) Surface temperature, (b) Liquid temperature and (c) Vapor temperature versus radius square for different components.

As the liquid droplet is in evaporation after 1.5s (figure 16.D), the two-phase flows and the heat transfer are still active. Hence, the vapor is spread in the whole area and its temperature is nearly equal to ambient temperature. The liquid and the liquid-free surface temperatures are higher and reach approximately the surface and the boiling temperatures of the hydrocarbon droplet respectively.

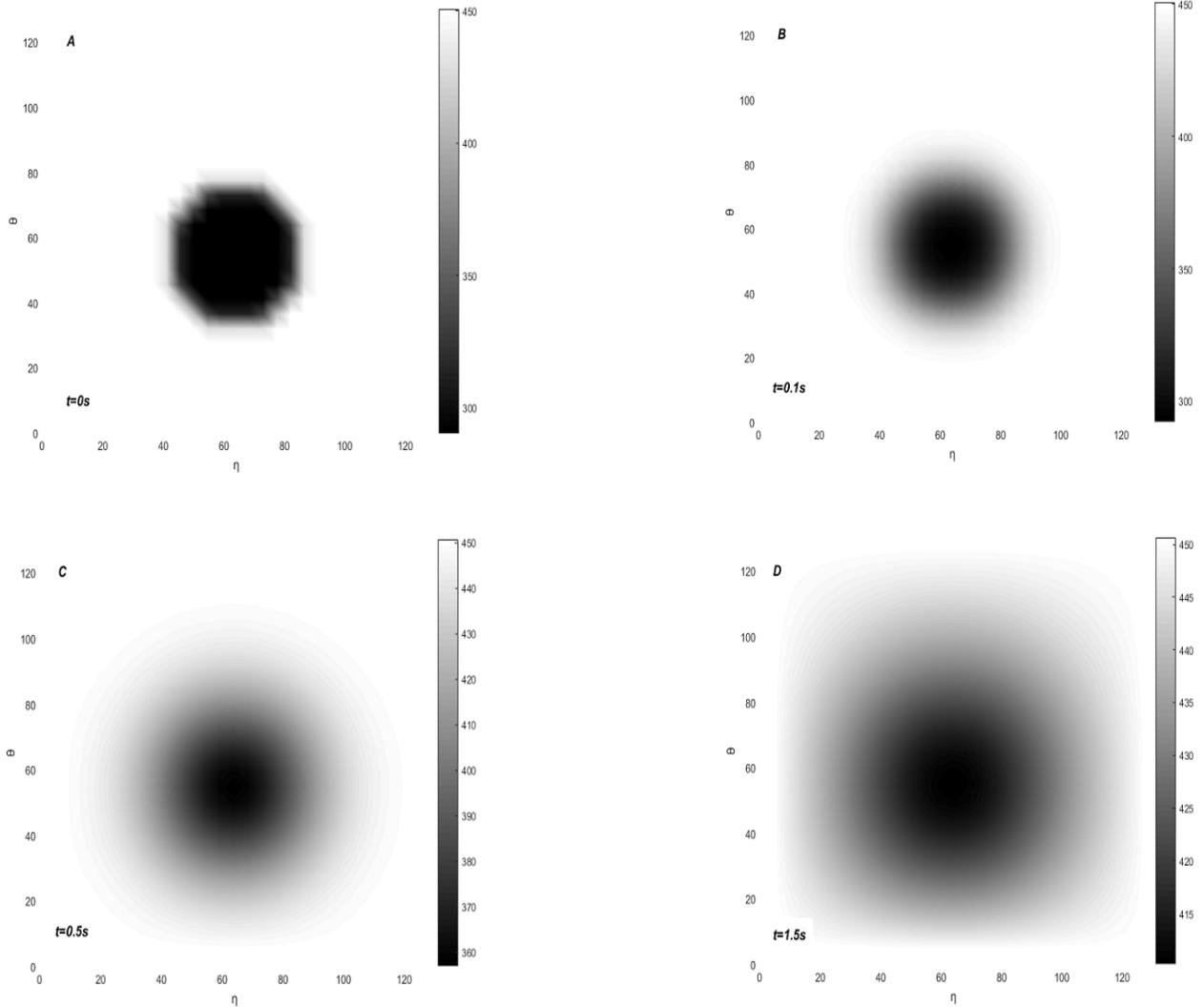

**Fig 16:** Isothermal sequence for an octane droplet in evaporation under the next conditions: $R_0 = 1.1\ mm$; $T_{L0} = 290.15\ K$; $T_{S0} = 293.15\ K$; $T_\infty = 450.63\ K$.

Other surface profiles, (figures 17 and 18), are realized and show the evaporation phenomenon of oxygenated biomass fuel droplets, methanol and ethanol, the heat and mass transfer, and the gradient temperature. At the initial time t=0s, figure 17.A represents the initial temperatures of the liquid droplet, its free surface (white contour), and its ambient medium.

At t=0.1s of the figure 17.B, the temperature gradient activates the heat and mass transfer and the temperature of the liquid methanol, and thus increases as the black contours show the convection factor inside the liquid phase.

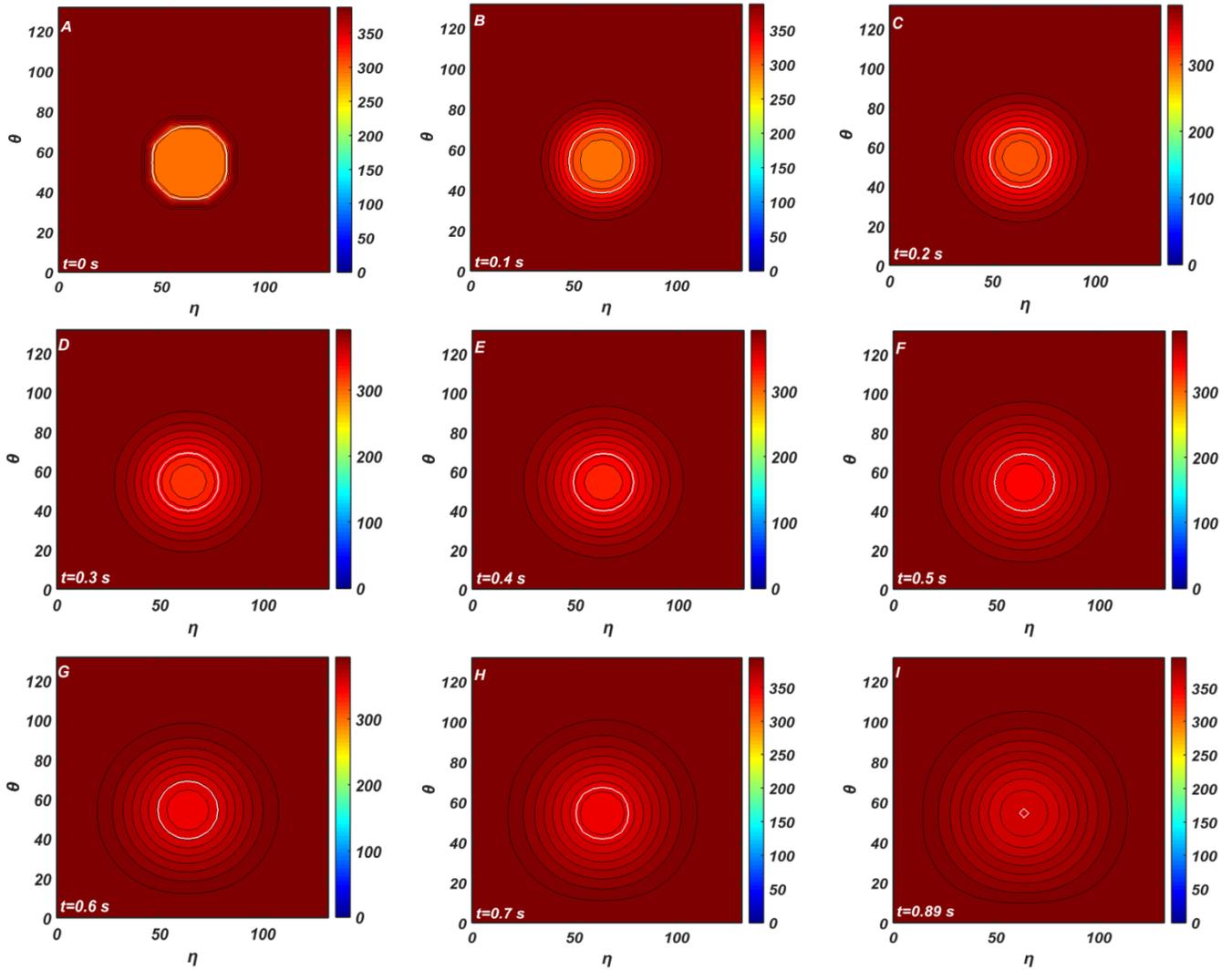

**Fig 17:** Isothermal and surface regression sequence for a methanol droplet in evaporation under the next conditions: $R_0 = 1.1 \ mm$; $T_{L0} = 292 \ K$; $T_{S0} = 295.43 \ K$; $T_\infty = 400 \ K$.

Moreover, the black contours, outside the liquid phase, show the formation of a circular vapor layer all around the droplet and the convection factor inside the vapor phase. This phenomenon is associated with the surface regression of the liquid droplet as the size of the white contour is becoming smaller. Thus, this observation continues in the next figures of the sequence to the total evaporation of the methanol liquid droplet after 0.89s, where the free surface is too small and the vapor is spread in the ambient medium in the form of a large circular cloud.

In figure 18, the sequence of the evaporation of the ethanol liquid droplet is based on the same phenomenon as the methanol liquid droplet and proves our observation of the fact as discussed in figure 17. The only difference is the time length of the phenomenon which is longer in figure 18 because the ethanol is less volatile than the methanol.

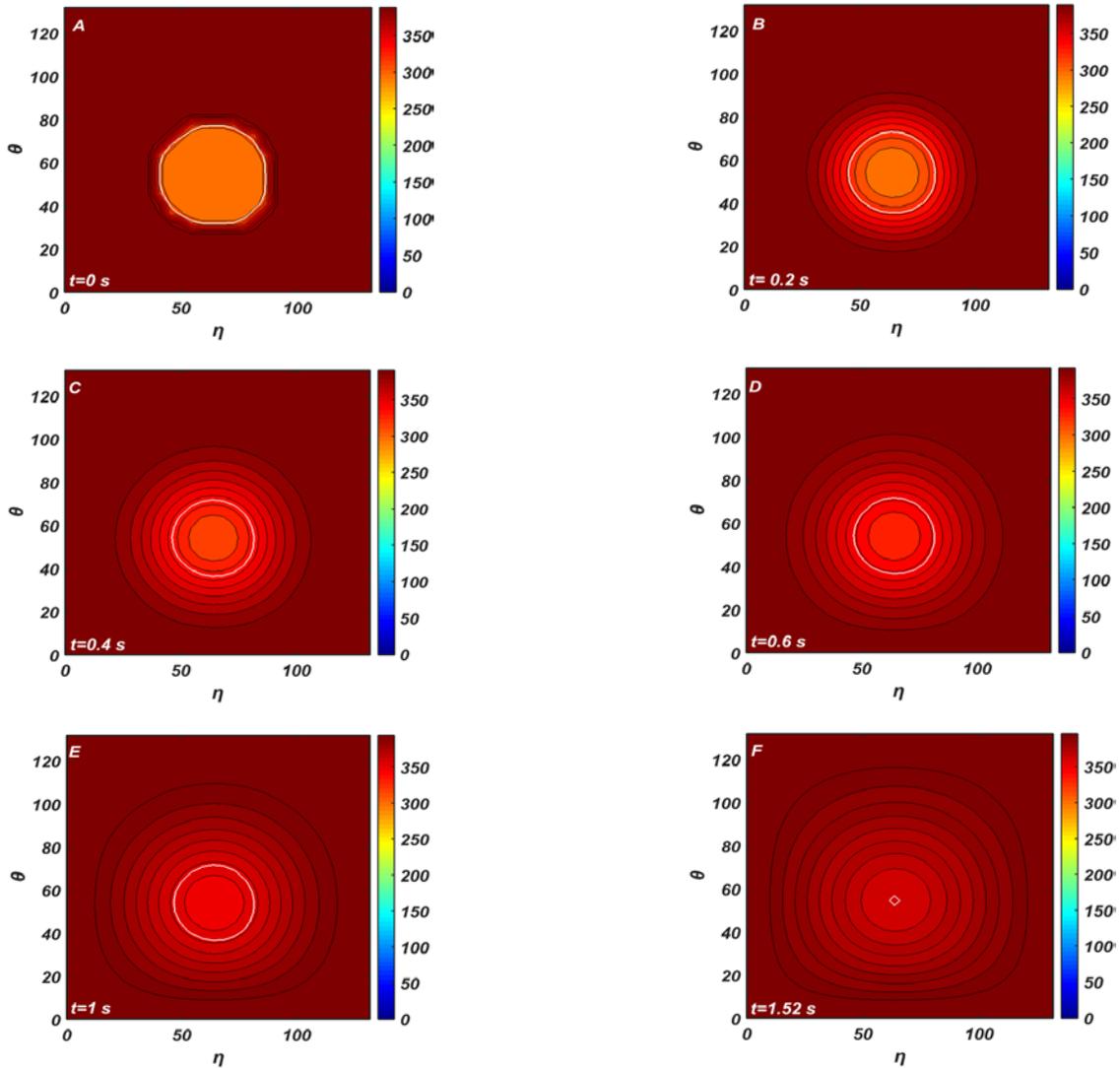

**Fig 18:** Isothermal and surface regression sequence for an ethanol droplet in evaporation under the next conditions: $R_0 = 1.3\ mm$; $T_{L0} = 292\ K$; $T_{S0} = 295.43\ K$; $T_\infty = 400\ K$

## IV. Conclusion

Navier-Stokes and energy equations of hydrocarbon liquid droplet are solved numerically using a semi-implicit method based on the finite volume scheme (*VOF*). A new estimation radius regression technique has been applied for liquid droplet evaporation, based on the *PLIC-VOF* method. This method tends to accurate radius calculation, to easy implementation, and not to be affected by the changing of the droplet's shape nor the temperature collecting and synchronization with the method for evaporation analysis. Thus, the calculation of the radius derived from the direct use of the color function properties is improved. Several linear and surface results are used to prove our numerical calculation. Our method shows good results comparing to the experimental ones. The comparison between radius, area, and mass regressions, volumetric mass flow, liquid temperatures, surface temperatures, and vapor temperatures of liquid hydrocarbon droplet evaporation is observed by taking into consideration the effect of ambient temperatures, radii, and components. Finally, using the time evolution of the color function, our radius regression estimation technique is based on instant detection of liquid volume fraction in each cell of the mesh.